\documentstyle[12pt,epsfig]{article}
\newlength{\dinwidth}
\newlength{\dinmargin}
\newcommand{\ba}{\begin{array}}
\newcommand{\ea}{\end{array}}
\newcommand{\beq}{\begin{equation}}
\newcommand{\eeq}{\end{equation}}
\newcommand{\bea}{\begin{eqnarray}}
\newcommand{\eea}{\end{eqnarray}}

\def\bce{\begin{center}}
\def\ece{\end{center}}

\def\nonu{\nonumber}

\def\pa{\partial}
\def\al{\alpha}

\def\de{\delta}

\def\ep{\epsilon}

\def\la{\lambda}
\def\La{\Lambda}

\def\si{\sigma}

\def\eps6{{\displaystyle \mathop{\epsilon}^{6}}{}}

\def\nab6{{\displaystyle \mathop{\nabla}^{6}}{}}

\begin{document}
\thispagestyle{empty}
\addtocounter{page}{-1}
\begin{flushright}
{\tt hep-th/0301011}\\
\end{flushright}
\vspace*{1.3cm}
\centerline{\Large \bf Supersymmetric $SO(N)/Sp(N)$ 
Gauge Theory} 
\vskip0.3cm
\centerline{\Large \bf from Matrix Model:Exact Mesonic Vacua}
\vspace*{1.5cm} 
\centerline{{\bf Changhyun Ahn}}
\vspace*{1.0cm}
\centerline{\it Department of Physics, 
Kyungpook National University, Taegu 702-701, Korea}
\vspace*{0.8cm}
\centerline{\tt ahn@knu.ac.kr}
\vskip2cm
\centerline{\bf Abstract}
\vspace*{0.5cm}

By performing the matrix integral over the tree level superpotential
of ${\cal N}=1 $ supersymmetric $SO(N)/Sp(N)$ gauge theories 
obtained from ${\cal N}=2$ SQCD by adding the mass term for the adjoint
scalar field,
the exact effective superpotential in terms of meson field  contains
the nonperturbative ADS superpotential as well as the classical tree
level superpotential. 
By completing the meson matrix integral with the help of saddle point
equation, we find the free energy contributions from matter part
in terms of glueball field, the adjoint field mass and quark mass.
By extremizing the effective superpotential 
with respect to the glueball field,
we analyze the vacuum structure and describe the behavior of
two limiting cases:zero limit of quark mass and  infinity limit of
adjoint field mass.
We also study the magnetic theory.

\vspace*{\fill}


\baselineskip=18pt
\newpage
\renewcommand{\theequation}{\arabic{section}\mbox{.}\arabic{equation}}

\section{Introduction}
\setcounter{equation}{0}

A technique for calculating the effective superpotential 
for the glueball
field in an ${\cal N}=1$ supersymmetric gauge theory
via a hermitian matrix integral over tree level superpotential was
proposed by Dijkgraaf and Vafa \cite{dv3,dv2,dv}.
Their work is based on mainly the ${\cal N}=1$ gauge theory coupled
to adjoint chiral fields with a general superpotential of a polynomial.
There are many related works \cite{cm}- \cite{feng3}
along the lines of \cite{dv3,dv2,dv}.
The matrix model calculation with flavors have been found in various 
places \cite{acfh1,mcg,suz,br1,dj,tachik,feng,fh,ohta,
bhr,hof,suz1,
seiberg,an,feng3}. 
In particular, in \cite{acfh1} the matrix integral
was completely calculated and infinite series in a perturbative expansion
was represented by a single analytic function. Based on this, the 
nonperturbative Affleck-Dine-Seiberg (ADS) superpotential was rederived in 
\cite{suz,br1}. On the other hand, there was a different approach to get ADS 
superpotential by including a matrix valued delta funcion, 
which is known as the Wishart integral \cite{dj}.
In \cite{fo,ino,ookouchi,ashoketal,feng1,jo,an},
there are some relevant 
works on the $SO(N)/Sp(N)$ gauge theories in the view point of matrix 
model \cite{dv3,dv2,dv}.    

In this paper,
we calculate the matrix path integral over tree level 
superpotential obtained from
${\cal N}=2$ SQCD by taking into account the 
mass term of adjoint scalar field. The adjoint field
mass $\mu$ is taken
to be much larger than the dynamical scale $\La_{N=2}$ 
of the ${\cal N}=2$
theory.  
In the gauge theory side,
one can consistently integrate out the adjoint scalar to obtain
a low-energy effective ${\cal N}=1$ superpotential.
This effective theory can be determined exactly using the information
of the low-energy degrees of freedom
of ${\cal N}=1$ $SO(N)$ gauge theories  which was discovered by
Intriligator and Seiberg \cite{is} and $Sp(N)$ gauge theories by
Intriligator and Pouliot \cite{ip}.
By inserting a matrix valued delta function initiated by Demasure and Janik 
\cite{dj} in our problem,
the exact effective superpotential in terms of meson field  possesses
the nonperturbative ADS superpotential \cite{is,ip} 
plus the classical tree
level superpotential. It is rather surprising to see this 
nonperturbative ADS superpotential within the Dijkgraaf-Vafa matrix
model.

By completing a  matrix integral over meson field, 
we find the free energy contribution from matter part
in terms of glueball field, the adjoint mass and quark mass.
By extremizing the effective superpotential obtained from
the contributions of free energies, 
with respect to the glueball field,
we analyze the mesonic vacuum structure.
In the gauge theory side, the number of ${\cal N}=1$ vacua
and the pattern of flavor symmetry breaking can be determined 
in the limit of zero mass of quark with $\mu$ and $\La_{N=2}$ fixed
while the infinity  limit of
adjoint field mass keeping the quark mass and  the scale of 
${\cal N}=1$ theory fixed corresponds to
the standard ${\cal N}=1$ theory without adjoint field.
For $U(N)$ gauge theory with $N_f$ flavors of quarks in the 
fundamental and anti-fundamental representations,
the exact mesonic vacua in
the matrix descriptions have been found in \cite{ohta}. 

\section{Matrix model description of supersymmetric $SO(N)$ theory }
\setcounter{equation}{0}
Let us
consider  an ${\cal N}=1$ supersymmetric $SO(N)$ gauge theory 
with $N_f$ flavors of quarks $Q^i_a(i=1, 2, \cdots, N_f(=2n_f), 
a=1, 2, \cdots, 
N)$ in the vector (fundamental)
representation ($N \geq 4, N_f \leq N-5$). 
The tree level superpotential of the theory is obtained from 
${\cal N}=2$ SQCD
by adding the mass $\mu$  for the adjoint scalar
$\Phi_{ab}$ belonging to the ${\cal N}=2$ vector
multiplet \cite{aps,as,hms,aot,carlinoetal}
\bea 
W_{tree}(\Phi, Q) = \frac{1}{2} \mu \mbox{Tr} \; \Phi^2 +
\sqrt{2} Q^i_a \Phi_{ab} Q^j_b J_{ij} + \frac{1}{2} m_{ij} Q^i_a Q^j_a
\label{tree1}
\eea
where 
 $J_{ij}$ is the symplectic metric 
$( {0 \atop -1 }{ 1 \atop 0}  ) \otimes {\bf 1}_{n_f \times n_f} $
used to raise and lower 
$SO(N)$ flavor indices ( $
{\bf 1}_{n_f \times n_f}$ is the $n_f \times n_f$ identity matrix )
and $m_{ij}$ is a quark mass matrix
$( { 0 \atop 1 }{  1 \atop 0 }  ) 
\otimes \mbox{diag} ( m_{1}, \cdots, m_{n_f} ) $.
The vacuum structure and phases of resulting ${\cal N}=1$
theories by integrating out the adjoint scalar was found in 
\cite{carlinoetal}. The $\mu=0$ theory (with ${\cal N}=2$ 
supersymmetry) has a $Z_{2N_c-N_f-4} \times SU(2)_R$ R-symmetry which
is broken to $Z_2 \times SU(2)_R$ by the vev of $\mbox{Tr} \Phi^2$.
On the other hand, in an 
${\cal N}=1$ theory with $\mu \neq 0$, the adjoint
mass explicitly breaks the R-symmetry down to $Z_2$. 

By manipulating this tree level superpotential as 
the potential for the matrix model,
we describe $SO(N)$ matrix model at large $N$ by 
replacing the gauge theory fields with matrices to
get the contributions to the free energies. Then the 
partition function can be written as
\bea
Z= \frac{1}{vol(SO(N))} 
\int  [d \Phi] [d Q] 
e^{-\frac{1}{g_s} W_{tree}(\Phi, Q)}
\label{Z} 
\eea
where the large $N$-limit behavior of the volume for $SO(N)$ gauge group
can be read off from  \cite{mac,ov,ashoketal}
\bea
\log vol(SO(N)) 
 =  -\frac{1}{4} N^2 \log \left( \frac{N}{2\pi e^{3/2}}
\right) + \cdots.
\nonu
\eea
In order to describe and study
the effective theory by the gauge invariant meson fields 
$M^{ij}=Q^i \cdot Q^j$, we make a change of variable in the superpotential
by using a matrix valued delta function defined in \cite{jn}
which is valid for the region $N_f \leq N-5$ 
along the line of \cite{dj} 
\bea
e^{-\frac{1}{g_s} W_{tree}(\Phi, Q)}=
\int [dM] \de(M^{ij}-Q^i \cdot Q^j) 
e^{-\frac{1}{g_s} W_{tree}(\hat{\Phi}, M)}
\label{delta1}
\eea
where the new superpotential in terms of  $\hat{\Phi}$
can be written as 
\bea
W_{tree}(\hat{\Phi}, M) = 
\frac{1}{2} \mu \mbox{Tr} \; \hat{\Phi}^2 -
\frac{1}{ 2\mu} \mbox{Tr} M J M J + \frac{1}{2} \mbox{Tr} mM, \qquad 
\hat{\Phi}_{ab} = \Phi_{ab} + \frac{\sqrt{2}}{\mu} Q^i_a Q^j_b J_{ij}.
\nonu
\eea
Now we substitute (\ref{delta1}) into (\ref{Z}) with matrix path integral
over $\hat{\Phi}$
\bea
Z  =  \frac{1}{vol(SO(N))}  
\int  [d \hat{\Phi}] [d Q] [dM] \de(M^{ij}-Q^i \cdot Q^j)
e^{-\frac{1}{g_s} W_{tree}(\hat{\Phi}, M)}.
\nonu 
\eea
We make a matrix gaussian integral over  $\hat{\Phi}$ (in the gauge theory
side this is equivalent to the statement that as the mass $\mu$ is
increased beyond the scale of the asymptotic free ${\cal N}=2$ theory
one can integrate out the adjoint field)
and get
\bea
Z & = & \frac{1}{vol(SO(N))}  \left(
\frac{4 \pi g_s}{\mu} \right)^{N^2/2} 
\int   [d Q] [dM] \de(M^{ij}-Q^i \cdot Q^j)
e^{-\frac{1}{g_s} W_{tree}( X)}
\nonu 
\eea
where the tree level superpotential depends on the meson fields $M^{ij}$ only
and
is given by
\bea
W_{tree}(M) = -
\frac{1}{ 2\mu} \mbox{Tr} M J M J + \frac{1}{2} \mbox{Tr} mM.
\label{wtreeso}
\eea

We execute a matrix integral over $Q$ and use the result of Wishart
random matrices \cite{dj,jn,an}
\bea
Z(S) & = & \frac{1}{vol(SO(N))}  \left(
\frac{4 \pi g_s}{\mu} \right)^{N^2/2} e^{-\frac{1}{2} N_f N \log \frac{N}{2}}
\int    [dM]  \left( \mbox{det} M \right)^{(N-N_f-1)/2}
e^{-\frac{1}{g_s} W_{tree}(M)} \nonu \\
& \equiv & \int    [dM]  Z(S,M).
\label{ZZ}
\eea
In the large $N$-limit we are interested in, the glueball 
field $S$ can be identified with $g_s N$ and
the size of $M$ depends on $N_f$ and the matrix integral over
$M$ does not contribute to the function of $S$.
Let us denote $Z(S,M)$ as the partition function before the integral
over $M$.
Then 
one can write
the log of the partition function as follows:
\bea
\log Z(S,M) 
&=& \frac{1}{4} N^2 \log \left(\frac{8\pi g_s^2 N}{ e^{3/2} \mu^2}\right)  
-\frac{1}{2} N_f 
N \log \left( \frac{N}{2} \right) + \frac{N}{2} 
 \log \mbox{det} M  -\frac{1}{g_s} W_{tree}(M) + \cdots \nonu \\
& = &  \frac{S^2}{4g_s^2} 
\log \left( \frac{8\pi g_s S}{ e^{3/2} \mu^2} \right)  -\frac{S N_f}{2g_s}  
\log \left( \frac{S}{2g_s} \right) + \frac{S}{2g_s} 
 \log \mbox{det} M
 -\frac{1}{g_s} W_{tree}(M) + \cdots \nonu \\
& \equiv & -\frac{1}{g_s^2} {\cal F}_2 -\frac{1}{g_s} {\cal F}_1.
\nonu
\eea

Then
the effective superpotential $W(S,M)$ for the glueball field $S$ by 
identifying the $1/g_s^2$ and $1/g_s$ terms can be 
computed as the derivative of the contribution to free energy 
${\cal F}_2 $ plus 
the contribution ${\cal F}_1$ from flavors
\cite{dv,acfh1,ino,ashoketal,jo}
\bea
W  & = &
\left(N_c-2\right) \frac{\pa {\cal F}_2}{\pa S} + {\cal F}_1
\nonu \\
& = & \frac{1}{2} \left(N_c-2 \right) 
\left[ S - S \log \left( \frac{S}{ \La^3} \right) \right]  -
\frac{S N_f}{2}  \left( 1- \log S \right) -
\frac{S}{2} \log \mbox{det}  M   + W_{tree}(M).
\label{b}
\eea
Solving the F-flatness condition $\pa_S W=0$ 
(minimizing $W(S,M)$ with respect to a glueball field $S$)
one gets
\bea
\hat{S} = \left( \frac{\La^{3\left(N_c-2\right)}} {\mbox{det} M}  
\right)^{1/\left(N_c-N_f-2\right)} 
e^{2\pi i k/\left(N_c-N_f-2\right)},  \qquad 
k=1, \cdots, \left(N_c-N_f-2\right)
\nonu
\eea
with the phase factor $e^{2\pi i k/\left(N_c-N_f-2\right)}$
reflecting the $(N_c-N_f-2)$ supersymmetric vacua. 
Then the exact superpotential $W(M)$ by plugging back 
$\hat{S}$ 
into (\ref{b}) leads to
\bea
W(M) & = &
\frac{1}{2} \left(N_c-N_f-2 \right) \hat{S}    + W_{tree}(M) \nonu \\
& = &  \frac{1}{2} \left(N_c  -N_f-2 \right) \ep_{N_c-N_f-2} \left( 
 \frac{\La^{3\left(N_c-2\right)}}{ \mbox{det}  M} 
\right)^{1/\left(N_c-N_f-2\right)} + W_{tree}(M) \nonu \\
&= &   \frac{1}{2} \left(N_c -N_f -2 \right) \ep_{N_c-N_f-2} \left( 
\frac{16 \La_{N_c,N_f}^{3\left(N_c-2\right)-N_f}}{\mbox{det} 
M }\right)^{1/\left(N_c-N_f-2\right)} + 
W_{tree}(M) \nonu \\
& =& W_{ADS}(M)+W_{tree}(M)
\label{eW1}
\eea
where
$\La^{3(N_c-2)}=  
16 \La_{N_c,N_f}^{3\left(N_c-2\right)-N_f} $ is the strong coupling
scale of the ${\cal N}=1$ theory obtained by decoupling
the adjoint field $\Phi$  and $\ep_{N_c-N_f-2} \equiv 
e^{2\pi i k/\left(N_c-N_f-2\right)}$ is the $(N_c-N_f-2)$-th root of
unity.
We have checked that the nonperturbative ADS superpotential
$W_{ADS}(M)$ \cite{is} was obtained from the large $N$ asymptotics of 
the constrained matrix integral measure 
in the context of Dijkgraaf-Vafa matrix
model.
In the low energy effective theory, the classical vacuum degeneracy was 
lifted by quantum effects which is represented by a dynamically 
generated superpotential for the light meson fields $M^{ij}$. 
The superpotential \cite{is} alone generated by gaugino condensation 
has no vacuum but by adding the $W_{tree}(M)$
to the superpotential $W_{ADS}(M)$,  
the theory has $(N_c-N_f-2)$ supersymmetric vacua.
If not all of the matter fields are massive, in the gauge theory side,
one can integrate  out massive quarks and get the effective superpotential
at low energy for the massless ones. 
It is the same form as the above $W_{ADS}(M)$ but with the scale replaced by
the low energy one.

In order to perform the matrix integral over $M$ in (\ref{ZZ}), 
the evaluation of this matrix integral is 
approximated around the solution of the saddle point equation 
(for $U(N)$ matrix model this approach was done in \cite{ohta}) 
\bea
-\frac{1}{\mu} \la_i + \frac{1}{2} m_i - \frac{S}{\la_i} =0 
\label{saddle}
\eea
where the matrix $M$ has the following form
$( {0 \atop B^T }{ B \atop 0}  )$ and $B = \mbox{diag}(\la_1, \cdots, 
\la_{n_f})$ \cite{carlinoetal}.
In the gauge theory side, this can be obtained by differentiating 
(\ref{eW1}) with respect to the meson fields $M$ and looking at
the equations of motion for the $\la_i$ \cite{carlinoetal}.
This has two solutions for each $\la_i$ as a function of
$S$ denoted by $\la_{i}^{(\pm)}$
\bea
\la_{i}^{(\pm)} = \mu m_i \left( \frac{1 \pm \sqrt{1-16 \al_i S }}{4}\right)
\equiv  \mu m_i f^{(\pm)}(\al_i S), \qquad \al_i =\frac{1}{\mu m_i^2}.
\label{lambda}
\eea
Let us remind that $m_i$'s are the component of a quark mass matrix  
$\label{mass}( { 0 \atop 1 }{  1 \atop 0 }  ) 
\otimes \mbox{diag} ( m_{1}, \cdots, m_{n_f} ) $.
By using the solution (\ref{saddle}) at the  saddle point and eliminating 
the quadratic part of meson field of the effective superpotential, 
one can read off the free energy in terms of glueball field $S$,
a quark mass $m_i$ and the adjoint field mass $\mu$ from the matter part
as follows:
\bea
{\cal F}_1(S, \al_i) = -S \sum_{i=1}^{n_f} \left(\frac{1}{2}-
\frac{1}{4 f^{(\pm)}(\al_i S)} -\log  f^{(\pm)}(\al_i S)  \right)
\nonu
\eea
where we also used the fact that the product of two roots of $\la_i$
behaves like 
\bea
\la_i^{(+)} \la_i^{(-)} =\mu S.
\nonu
\eea

It is ready to write the effective superpotential from 
the complete matrix integration 
and the free energy due to the matter part above in terms of guleball 
field $S$:
\bea
W(S,\mu,m_i,\La) = S \left[ \frac{1}{2} \left(N_c-2 \right) \left( 
 1 -  \log  \frac{S}{ \La^3}
\right)-
\sum_{i=1}^{n_f} \left(\frac{1}{2}-
\frac{1}{4 f^{(\pm)}(\al_i S)} -\log  f^{(\pm)}(\al_i S)  \right) \right].
\nonu
\eea
By differentiating this with respect to $S$ and using the explicit expression
for $f^{(\pm)}(\al_i S)$ in (\ref{lambda}) we get
\bea
\left( \frac{S}{\La^3} \right)^{\frac{N_c-2}{2}} & = &
\prod_{i=1}^{n_f}  f^{(\pm)}(\al_i S)  
\nonu \\
& = & \frac{1}{2^{2n_f}} 
\prod_{i=1}^r \left( 1 + \sqrt{1-16 \al_i S} \right)
\prod_{i'=r+1}^{n_f} \left( 1 - \sqrt{1-16 \al_i S} \right)
\label{Svacua}
\eea
where the index $i$ runs for the positive sign among the roots
of $\la_{i}^{(\pm)}$ and the index $i'$
does for the negative sign respectively. 
Here we used 
${f^{(\pm)}}(\al_i S)^2-\frac{1}{2}f^{(\pm)}(\al_i S) + \al_i S=0$.
This relation (\ref{Svacua}) 
is exactly the same as the defining equation of $X$ in the
notation of \cite{carlinoetal}.
Although one cannot solve exactly, one determines
the number of distinct solutions (or the number of distinct vacua)
and symmetry breaking patterns in certain limiting regimes. 
In the $m_i \rightarrow 0$ limit ($m_i <<  \La_{N=2} << \mu $),
one can ignore the second term in (\ref{saddle}) and from 
(\ref{Svacua}) one gets
\bea
\la_i = \la^{(\pm)} \sim \pm \sqrt{-\mu S}, \qquad
S^{N_c-n_f-2} \sim \left( \mu \La_{N=2}^2 \right)^{N_c-n_f-2}.
\nonu
\eea
Here the strong coupling scale of the ${\cal N}=1$ theory is related to
the dynamical ${\cal N}=2$ scale $\La_{N=2}$  by one loop matching
condition through
\bea
\mu^{N_c-2} \La_{N=2}^{2(N_c-2-n_f)} = \La_{N_c,N_f}^{3(N_c-2)-N_f}.
\nonu
\eea 
Then there exist $(N_c-n_f-2)({ n_f \atop r })$ distinct solutions 
(that is, the expression of (\ref{Svacua}) is a polynomial
of degree $(N_c-n_f-2)$ in $S$ and the number of ways for choosing
$r$ positive signature is given by $({ n_f \atop r })$) 
and the total number of ${\cal N}=1$ vacua is 
given by
\bea
\# = (N_c-n_f-2) \sum_{r=0}^{n_f} ({ n_f \atop r }) =
 \left( N_c-n_f-2\right) 2^{n_f}
\nonu
\eea
which coincides with the number of the semi-classical vacua 
\cite{carlinoetal}. The quantum
corrected effective action shows a spontaneous breakdown of the global
$Sp(2n_f)$ symmetry into $U(n_f)$ since
some of the meson vevs remain non-zero in this limit.
The proof for this was given in \cite{carlinoetal}.
When the quark masses are nonvanishing, 
the unbroken flavor symmetry 
is $U(r) \times U(n_f-r)$ which agrees with the classical description.

The solutions of the full nonlinear coupled equations 
(\ref{lambda}) and (\ref{Svacua}) can be classified according to
the number of $\la_i$'s.
For large quark masses ($\La_{N=2} << m_i << \mu$) assuming that
$S << \mu m_i^2$,  
the multiplicity of solutions gives
$(N_c-2r -2)$ and  there are $({ n_f \atop r })$ ways of selecting
a vacuum configuration with $r$ nonzero $\la^{(+)}$ and
the number of vacua with  $U(r) \times U(n_f-r)$ symmetry
is $(N_c-2r-2)({ n_f \atop r })$ which is the same as the 
semi-classical results.  
In particular, $r=0$ vacuum will lead to 
\bea
S^{N_c-2} \sim \mu^{N_c-2} \La^{2(N_c-n_f-2)}_{N=2} \mbox{det} m  \sim
\La_{N_c,N_f}^{3(N_c-2)-N_f} \mbox{det} m
\nonu
\eea
which was observed in \cite{an} in the description of 
Dijkgraaf-Vafa matrix model.

As the number of flavors is increased,
in the IR theory of electric theory, the magnetic theory
is described by an $SO(\widetilde{N}=N_f-N+4)$ gauge theory
($N_f > N \geq 4 $) 
with $N_f$ flavors of dual quarks $q^i_a(i=1, \cdots, N_f(=2n_f), 
a=1, \cdots, 
\widetilde{N})$ and the additional
gauge singlet fields $M^{ij}$ which is an elementary field of dimension 1
at the UV-fixed point \cite{is}.
The matter field variables in the magnetic theory
are the original electric variables $M^{ij}$ and magnetic quarks
$q_i$ with the superpotential together with mass term
\bea
W = \frac{1}{2 \kappa} M^{ij} q_i \cdot q_j + W_{tree}(M)
\nonu
\eea
where $W_{tree}(M)$ is given in (\ref{wtreeso}).
The partition function can be written 
as follows:
\bea
Z & = & \frac{1}{vol(SO(\widetilde{N}))}  
 \int  [d q]  [dM] e^{-\frac{1}{g_s} \left(   
 \frac{1}{ 2\kappa}  M^{ij} q_i \cdot q_j +
W_{tree}(M) \right)}. 
\nonu
\eea
After calculating a gaussian matrix integral over $q$, 
the effective superpotential from the
log of partition function
can be expressed as by using the method in \cite{an} in dual gauge
theory 
\bea
W(M) & = & \frac{1}{2} \left(\widetilde{N}_c-2 \right) 
\left( S - S \log \frac{S}{ 
{\widetilde{\La}}^3} \right)   
-
\frac{S N_f}{2} \log {\widetilde{\La}} + \frac{1}{2} S 
\log \mbox{det} \left( \frac{M}{\kappa} \right) +W_{tree}(M) 
\nonu \\
& = &  \frac{1}{2} \left(\widetilde{N}_c-2 \right) 
\left( S-  S \log  \frac{S} 
{\left( \widetilde{\La}^{3\left(\widetilde{N}_c-2\right)-
N_f} \mbox{det}  \left( \frac{M}{\kappa} \right) \right)^{1/
\left(\widetilde{N}_c-2\right)}} \right) + W_{tree}(M). 
\label{W12}
\eea
Solving the F-flatness condition $\pa_S W=0$
one gets, with the phase factor $e^{2\pi i k/\left(
\widetilde{N}_c-2\right)}, 
k=1, \cdots, \left(\widetilde{N}_c-2\right)$
reflecting the $(\widetilde{N}_c-2)$ supersymmetric vacua,
\bea
\hat{S}=\left( 2^{12} \widetilde{\La}^{3\left(\widetilde{N}_c-2\right)-
N_f}_{N_f-N_c+4,N_f} \mbox{det} \left( \frac{M}{\kappa} \right) 
 \right)^{1/\left(\widetilde{N}_c-2\right)} \ep_{
\widetilde{N}_c-2}, \qquad
\widetilde{\La}^{3\left(\widetilde{N}_c-2\right)-
N_f}
= 2^{12} \widetilde{\La}^{3\left(\widetilde{N}_c-2\right)-
N_f}_{N_f-N_c+4,N_f}.
\nonu
\eea

Then the exact superpotential by plugging this $\hat{S}$ into 
(\ref{W12}) leads to
\bea
W & = & 
\frac{1}{2} \left(\widetilde{N}_c-2 \right) \hat{S} +W_{tree}(M)
\nonu \\
& = & \frac{1}{2} \left(\widetilde{N}_c  -2 \right) \ep_{
\widetilde{N}_c-2} \left( 
\frac{ 2^{12} \widetilde{\La}_{N_f-N_c+4,N_f}^
{3\left(\widetilde{N}_c-2\right)-N_f} \mbox{det} M }
{ \kappa^{N_f}} \right)^{1/\left(\widetilde{N}_c-2\right)}+W_{tree}(M)
\nonu \\
& = & \frac{1}{2} \left(-N_c +N_f+ 2 \right)
\ep_{-N_c+N_f+2}
\left( 
\frac{ 2^{12} \widetilde{\La}_{N_f-N_c+4,N_f}^{3\left(N_f-N_c+2\right)-N_f}
\mbox{det} M}
{ \kappa^{N_f} }\right)^{1/\left(-N_c+N_f+2\right)}+W_{tree}(M)
\nonu \\
& = & 
 \frac{1}{2} \left(N_c -N_f -2 \right) \ep_{N_c-N_f-2} \left( 
\frac{16  \La_{N_c,N_f}^{3\left(N_c-2\right)-N_f}}{\mbox{det} M}  \right)^
{1/\left(N_c-N_f-2\right)}+W_{tree}(M)
\nonu \\
& = & 
W_{ADS}(M)+W_{tree}(M)
\label{W22}
\eea
which is exactly the same as the one in (\ref{eW1}). 
Here we used the fact that 
$\widetilde{N}_c = N_f -N_c +4$ and
$\ep_{
\widetilde{N}_c-2}=\ep_{-N_c+N_f+2}=\ep_{N_c-N_f-2}$.
The scale of the magnetic theory in the gauge theory side was 
related to that of 
electric theory by \cite{is}
\bea
2^8 \La_{N_c,N_f}^{3\left(N_c-2\right)-N_f} 
\widetilde{\La}_{N_f-N_c+4,N_f}^{3\left(N_f-N_c+2\right)-N_f} = 
\left(-1\right)^{N_f-N_c} 
\kappa^{N_f}
\nonu
\eea
where the normalization factor $1/2^8$ was chosen to get
the consistent low energy behavior under large mass deformation and along the 
flat directions.
Note that the factor $(-1)^{\frac{(N_f-N_c)}{(-N_c+N_f+2)}}=-1$ 
in (\ref{W22}) is cancelled exactly
by the overall $-1$ factor.

\section{Matrix model description of supersymmetric $Sp(N)$ theory }
\setcounter{equation}{0}

In this section, we continue to study the matrix model  
for the symplectic group $Sp(N)$.
Let us
consider  an ${\cal N}=1$ supersymmetric $Sp(N=2n)$ gauge theory 
with $N_f(=2n_f)$ flavors of quarks $Q^i_a(i=1, 2, 
\cdots, N_f, a=1, 2, \cdots, 
N)$ in the fundamental
representation ($N_f \leq N$). 
The tree level superpotential of the theory is obtained from 
${\cal N}=2$ SQCD
by adding the mass $\mu$  for the adjoint scalar
$\Phi_{ab}$ belonging to the ${\cal N}=2$ vector
multiplet \cite{aps,as,hms,aot1,carlinoetal1} 
\bea
W_{tree}(\Phi, Q) =  \mu \mbox{Tr} \; \Phi^2 +
\frac{1}{\sqrt{2}} Q^i_a \Phi^a_b Q^i_c J^{bc} + \frac{1}{2} m_{ij} Q^i_a 
Q^j_b J^{ab}
\label{tree}
\eea
where 
 $J_{ab}$ is the symplectic metric 
$( {0 \atop -1 }{ 1 \atop 0}  ) \otimes {\bf 1}_{n \times n} $
and $m_{ij}$ is a quark mass matrix
$( { 0 \atop 1 }{  -1 \atop 0 }  ) 
\otimes \mbox{diag} ( m_{1}, \cdots, m_{n_f} ) $.
The vacuum structure and phase and flavor symmetry breaking pattern
of this ${\cal N}=1$ theory was studied in \cite{carlinoetal1}. 

By manipulating this tree level superpotential as 
the potential for the matrix model,
we describe $Sp(N)$ matrix model at large $N$ by 
replacing the gauge theory fields with corresponding matrices to
get the contributions to the free energies. Then the 
partition function can be written as
\bea
Z= \frac{1}{vol(Sp(N))} 
\int  [d \Phi] [d Q] 
e^{-\frac{1}{g_s} W_{tree}(\Phi, Q)}
\label{Z1} 
\eea
where the large $N$-limit behavior of the volume for $Sp(N)$ gauge group   
can be read off from  \cite{mac,ov,ashoketal}
\bea
\log vol(Sp(N)) 
 =  -\frac{1}{4} N^2 \log \left( \frac{N}{2\pi e^{3/2}} \right) + \cdots.
\nonu
\eea
In order to describe the effective theory by the meson fields 
$M^{ij}=Q^i \cdot Q^j=Q_{ia}Q_{jb}J^{ab}=-M^{ji}$ which is 
antisymmetric, 
we make a change of variable in the superpotential
by using a matrix valued delta function \cite{dj} 
\bea
e^{-\frac{1}{g_s} W_{tree}(\Phi, Q)}=
\int [dM] \de(M^{ij}-Q^i \cdot Q^j) 
e^{-\frac{1}{g_s} W_{tree}(\hat{\Phi}, M)}
\label{delta}
\eea
where the new superpotential in terms of  $\hat{\Phi}$ 
can be written as 
\bea
W_{tree}(\hat{\Phi}, M) = 
\mu \mbox{Tr} \; \hat{\Phi}^2 -
\frac{1}{ 8\mu} \mbox{Tr} M J M J - \frac{1}{2} \mbox{Tr} mM, \qquad 
\hat{\Phi}^a_b = \Phi^a_b + \frac{1}{2\sqrt{2} \mu} J^{ac} Q^i_c Q^i_b.
\nonu
\eea

Now we substitute (\ref{delta}) into (\ref{Z1}) with matrix path integral
over $\hat{\Phi}$
\bea
Z  =  \frac{1}{vol(Sp(N))}  
\int  [d \hat{\Phi}] [d Q] [dM] \de(M^{ij}-Q^i \cdot Q^j)
e^{-\frac{1}{g_s} W_{tree}(\hat{\Phi}, M)}.
\nonu 
\eea
We make a matrix gaussian integral over  $\hat{\Phi}$
\bea
Z & = & \frac{1}{vol(Sp(N))}  \left(
\frac{2\pi g_s}{\mu} \right)^{N^2/2} 
\int   [d Q] [dM] \de(M^{ij}-Q^i \cdot Q^j)
e^{-\frac{1}{g_s} W_{tree}(M)}
\nonu 
\eea
where the tree level superpotential depends on the meson fields $M^{ij}$ and
is given by
\bea
W_{tree}(M) = -
\frac{1}{ 8\mu} \mbox{Tr} M J M J - \frac{1}{2} \mbox{Tr} mM.
\label{treeSp}
\eea
We execute a matrix integral over $Q$ and use the result of Wishart
random matrices \cite{dj,jn,an}
\bea
Z(S) & = & \frac{1}{vol(Sp(N))}  \left(
\frac{2\pi g_s}{\mu} \right)^{N^2/2} e^{-\frac{1}{2} N_f N \log \frac{N}{2}}
\int    [dM]  \left( \mbox{det} M \right)^{(N-N_f-1)/2}
e^{-\frac{1}{g_s} W_{tree}(M)}
 \nonu \\
& \equiv & \int    [dM]  Z(S,M).
\nonu
\eea
In the large $N$-limit we are interested in, the glueball 
field $S$ can be identified with $g_s N$ and
the size of $M$ depends on $N_f$ and the matrix integral over
$M$ does not contribute to the function of $S$.
Let us put $Z(S,M)$ as the partition function before the integral
over $M$.
Then 
one can write
the log of the partition function as follows:
\bea
\log Z(S,M) 
&=&   \frac{1}{4} N^2 \log \left(\frac{8\pi g_s^2 N}{ e^{3/2} \mu^2}\right) 
 -\frac{1}{2} N_f 
N \log \left( \frac{N}{2} \right) + \frac{N}{2} 
 \log \mbox{det} M  -\frac{1}{g_s} W_{tree}(M) + \cdots \nonu \\
& = &  \frac{S^2}{4g_s^2} 
\log \left( \frac{8\pi g_s S}{ e^{3/2} \mu^2} \right)  -\frac{S N_f}{2g_s}  
\log \left( \frac{S}{2g_s} \right) + \frac{S}{2g_s} 
 \log \mbox{det} M
 -\frac{1}{g_s} W_{tree}(M) + \cdots \nonu \\
& \equiv & -\frac{1}{g_s^2} {\cal F}_2 -\frac{1}{g_s} {\cal F}_1.
\nonu
\eea

Then
the effective superpotential $W(S,M)$ for the glueball field $S$ can be 
computed as the derivative of the contribution to free energy 
${\cal F}_2$ plus 
the contribution from flavors
\cite{dv,acfh1,ino,ashoketal}
\bea
W  & = &
\left(N_c+2\right) \frac{\pa {\cal F}_2}{\pa S} + {\cal F}_1
\nonu \\
& = & \frac{1}{2} \left(N_c+2 \right) 
\left[ S - S \log \left( \frac{S}{ \La^3} \right) \right]  -
\frac{S N_f}{2}  \left( 1- \log S \right) -
\frac{S}{2} \log \mbox{det}  M   + W_{tree}(M).
\label{WW}
\eea
Solving the F-flatness condition $\pa_S W=0$ 
(minimizing $W(S,M)$ with respect to a glueball field $S$)
one gets
\bea
\hat{S} = \left( \frac{\La^{3\left(n_c+1\right)}} {\mbox{Pf} M}  
\right)^{1/\left(n_c-n_f+1\right)} 
e^{2\pi i k/\left(n_c-n_f+1\right)},  \qquad 
k=1, \cdots, \left(n_c-n_f+1\right)
\nonu
\eea
with the phase factor $e^{2\pi i k/\left(n_c-n_f+1\right)}$
reflecting the $(n_c-n_f+1)$ supersymmetric vacua. 
Here $\mbox{Pf} M$ is the Pfaffian of the antisymmetric matrix $M$. 
Then the exact superpotential by plugging $\hat{S}$ into (\ref{WW}) leads to
\bea
W(M) & = &   \frac{1}{2} \left(N_c-N_f+2 \right) \hat{S} + 
W_{tree}(M) \nonu \\
& =& \left(n_c  -n_f+1 \right) \ep_{n_c-n_f+1} \left( 
\frac{\La^{3\left(n_c+1\right)}}{ \mbox{Pf}  M} 
\right)^{1/\left(n_c-n_f+1\right)} + W_{tree}(M) 
\nonu \\
&= &   \left(n_c -n_f +1 \right) \ep_{n_c-n_f+1} \left( 
\frac{2^{n_c-1} \La_{n_c,n_f}^{3\left(n_c+1\right)-n_f}}{\mbox{Pf} 
M }\right)^{1/\left(n_c-n_f+1\right)} + 
W_{tree}(M) 
\nonu \\
& =& W_{ADS}(M)+W_{tree}(M)
\label{eW}
\eea
where
$\La^{3(n_c+1)}=  
2^{n_c-1} \La_{n_c,n_f}^{3\left(n_c+1\right)-n_f} $.
The classical vacuum degeneracy was 
lifted by quantum effects which is represented by a dynamically 
generated superpotential for the light meson fields $M^{ij}$. 
The superpotential \cite{ip} generated by gaugino condensation 
has no vacuum but by adding the (\ref{treeSp})
to the superpotential $W_{ADS}(M)$,  
the theory has $(n_c-n_f+1)$ supersymmetric vacua.

In order to perform the matrix integral over $M$, it is 
approximated around the solution of the saddle point equation 
(for $U(N)$ matrix model this approach was done in \cite{ohta}) 
\bea
\frac{1}{2\mu} \la_i -  m_i - \frac{S}{\la_i} =0 
\label{saddleSp}
\eea
where the matrix $M$ has the following form
$i \si_2 \otimes \mbox{diag}(\la_1, \cdots, 
\la_{n_f})$.
This has two solutions for each $\la_i$
\bea
\la_{i}^{(\pm)} = \mu m_i \left( 1 \pm \sqrt{1+2 \al_i S }\right)
\equiv  \mu m_i f^{(\pm)}(\al_i S), \qquad \al_i =\frac{1}{\mu m_i^2}.
\nonu
\eea
By using the solution (\ref{saddleSp}) at the  saddle point and eliminating 
the quadratic part of meson field of the effective superpotential, 
together with the fact that
\bea
\la_i^{(+)} \la_i^{(-)} =-2\mu S,
\label{lambda1}
\eea
one gets the free energy in terms of glueball field, quark mass and
adjoint field mass from the matter part
\bea
{\cal F}_1(S, \al_i) = -S \sum_{i=1}^{n_f} \left(\frac{1}{2}-
\frac{1}{ f^{(\pm)}(\al_i S)} -\log  f^{(\pm)}(\al_i S)  \right).
\nonu
\eea

The effective superpotential from 
the complete matrix integration 
and the free energy due to the matter part above in terms of glueball 
field $S$ is
\bea
W(S,\mu,m_i,\La) = S \left[ \frac{1}{2} \left(N_c+2 \right) \left( 
 1 -  \log  \frac{S}{ \La^3}
\right)-
\sum_{i=1}^{n_f} \left(\frac{1}{2}-
\frac{1}{ f^{(\pm)}(\al_i S)} -\log  f^{(\pm)}(\al_i S)  \right) \right].
\nonu
\eea
By differentiating this with respect to $S$ we get
\bea
\left( \frac{S}{\La^3} \right)^{n_c+1} & = &
\prod_{i=1}^{n_f}  f^{(\pm)}(\al_i S)  
\nonu \\
& = &
\prod_{i=1}^r \left( 1 - \sqrt{1+2 \al_i S} \right)
\prod_{i'=r+1}^{n_f} \left( 1 + \sqrt{1+2 \al_i S} \right)
\nonu
\eea
where we choose 
$r$ negative signs and $(n_f-r)$ positive signs in the roots of 
$\la_i$.
Here we used 
${f^{(\pm)}}(\al_i S)^2-2f^{(\pm)}(\al_i S) -2 \al_i S=0$
obtained from (\ref{saddleSp}) easily.
As in the case of previous section, this relation corresponds to
the defining equation of $X$ in \cite{carlinoetal1}.
The distinct vacua can be determined in certain limiting regimes.
In the massless quark limit ($m_i \rightarrow 0$) 
with fixed $\mu$ and $\La_{N=2}$, 
there are $(2n_c-n_f+2)$ solutions for $S$ by ignoring
the second term in (\ref{saddleSp}) and 
there is $2^{n_f}$ possibilities for the signs among $\la_i$'s
\bea
\la_i = \la^{(\pm)} \sim \pm \sqrt{2 \mu S}, \qquad
S^{2n_c-n_f+2} \sim \left( \mu \La_{N=2}^2 \right)^{2n_c-n_f+2}.
\nonu
\eea
where 
the strong coupling scale of the ${\cal N}=1$ theory is related to
the dynamical ${\cal N}=2$ scale $\La_{N=2}$  through
\bea
\mu^{2n_c+2} \La_{N=2}^{2(2n_c+2-n_f)} = 
\La_{n_c,n_f}^{2(3(n_c+1)-n_f)}.
\nonu
\eea 

Moreover, for a particular phase of $S$ with even or odd solutions, 
the number of 
minus signs among $\la_i$ must be even or odd, respectively. 
This will reduce to the choice of the signs by half 
\cite{carlinoetal1}. 
Therefore there exists 
\bea
 \left( 2n_c-n_f+2\right) 2^{n_f-1}
\nonu
\eea
vacua which is consistent with the semi-classical result. 

For $\mu \rightarrow \infty$ limit which is different from previous 
limit,  some of the $\la_i$'s of (\ref{lambda1}) are of order of
$\mu$ while others are much smaller. For large $\la_i$'s there 
exists the relation $\la_i \sim 2m_i \mu$ and the smaller $\la_i$'s
can be found by substituting the large $\la_i$'s into the defining
equation (\ref{saddleSp}). 
According to the analysis of 
\cite{carlinoetal1}, the total number of the vacua agrees with 
the one of the classical vacua.    
We can see the standard supersymmetric vacua of the theories 
without adjoint field corresponding to $r=0$ vacua
\bea
S^{n_c+1} \sim \mu^{n_c+1} \La^{2(n_c+1)-n_f}_{N=2} \mbox{Pf} m
\sim \La_{n_c,n_f}^{3(n_c+1)-n_f} \mbox{Pf} m  
\nonu
\eea
which can be interpreted similarly as the one in \cite{an}.
For $N_f=N_c+2$, it is evident that ADS superpotential vanishes
and the classical moduli space of vacua is changed quantum mechanically.
The large $\mu$ theory develops a quantum modified
constraint \cite{ip}
\bea
\mbox{Pf} M = 2^{n_c-1} \La_{n_c,n_c+1}^{2(n_c+1)}.
\nonu
\eea   
By introducing a Lagrange multiplier $Y$
in order to impose this constaint
the effective  superpotential will be 
\bea
W = Y \left( \mbox{Pf} M - 2^{n_c-1} \La_{n_c,n_c+1}^{2(n_c+1)}
\right) + W_{tree}(M).
\nonu 
\eea
One can check that the total number of vacua will be consistent with
the semi-classical result.
When $N_f=N_c+4$, the large $\mu$ theory develops a superpotential
\bea
W = -\frac{\mbox{Pf} M}{
 2^{n_c-1} \La_{n_c,n_c+2}^{2(n_c+1)}}
 + W_{tree}(M).
\nonu
\eea 

In the IR theory of electric theory, the magnetic theory
is described by an $Sp(\widetilde{N}=N_f-N-4)$ gauge theory
($N_f > N+4$) 
with $N_f$ flavors of dual quarks $q^i_a(i=1, \cdots, N_f(=2n_f), 
a=1, \cdots, 
\widetilde{N})$ and the additional
gauge singlet fields $M^{ij}$ which is an elementary field of dimension 1
at the UV-fixed point \cite{ip}.
The matter field variables in the magnetic theory
are the original electric variables $M^{ij}$ and magnetic quarks
$q_i$ with the superpotential together with mass term
\bea
W = \frac{1}{ \kappa} M^{ij} q_i \cdot q_j + W_{tree}(M).
\nonu
\eea
where the scale $\kappa$ is the matching scale between the electric
and magnetic gauge couplings and $W_{tree}(M)$ is given in
(\ref{treeSp}).  
The partition function can be written 
as follows:
\bea
Z & = & \frac{1}{vol(Sp(\widetilde{N}))}  
 \int  [d q]  [dM] e^{-\frac{1}{g_s} \left(   
 \frac{1}{ \kappa}  M^{ij} q_i \cdot q_j +
W_{tree}(M) \right)}. 
\nonu
\eea

The effective superpotential from the
log of partition function
after calculating a gaussian matrix integral over $q$ 
can be expressed as
\bea
W & = & \frac{1}{2} \left(\widetilde{N}_c+2 \right) 
\left( S - S \log \frac{S}{ 
{\widetilde{\La}}^3} \right)   
-
\frac{S N_f}{2} \log {\widetilde{\La}} + \frac{1}{2} S 
\log \mbox{det} \left( \frac{2M}{\kappa} \right) +W_{tree}(M) 
\nonu \\
& = &  \frac{1}{2} \left(\widetilde{N}_c+2 \right) 
\left( S-  S \log  \frac{S} 
{\left( \widetilde{\La}^{3\left(\widetilde{N}_c+2\right)-
N_f} \mbox{det}  \left( \frac{2M}{\kappa} \right) \right)^{1/
\left(\widetilde{N}_c+2\right)}} \right)+W_{tree}(M) 
\nonu \\
& = &   \left(\widetilde{n}_c+1 \right) 
\left( S-  S \log  \frac{S} 
{\left( \widetilde{\La}^{3\left(\widetilde{n}_c+1\right)-
n_f} \mbox{Pf}  \left( \frac{2M}{\kappa} \right) \right)^{1/
\left(\widetilde{n}_c+1\right)}} \right) + W_{tree}(M) 
\label{a}
\eea
where $\widetilde{n}_c = \widetilde{N}_c/2$.
Solving the F-flatness condition $\pa_S W=0$
one gets, with the phase factor $e^{2\pi i k/\left(
\widetilde{n}_c+1\right)}, 
k=1, \cdots, \left(\widetilde{n}_c+1\right)$
reflecting the $(\widetilde{n}_c+1)$ supersymmetric vacua,
\bea
\hat{S}=\left( 2^{\widetilde{n}_c-1} 
\widetilde{\La}^{3\left(\widetilde{n}_c+1\right)-
n_f}_{n_f-n_c-2,n_f} \mbox{Pf}  \left( \frac{2M}{\kappa}
\right) 
 \right)^{1/\left(\widetilde{n}_c+1\right)} \ep_{
\widetilde{n}_c+1}, \qquad
\widetilde{\La}^{3\left(\widetilde{n}_c+1\right)-
n_f}
= 2^{\widetilde{n}_c-1} 
\widetilde{\La}^{3\left(\widetilde{n}_c+1\right)-
n_f}_{n_f-n_c-2,n_f}.
\nonu
\eea

Then the exact superpotential by plugging 
this $\hat{S}$ into (\ref{a}) leads to
\bea
W & = &  
\frac{1}{2} \left(\widetilde{N}_c+2 \right) \hat{S}
+ W_{tree}(M)
\nonu \\
& = & \left(\widetilde{n}_c  +1 \right) \ep_{
\widetilde{n}_c+1} \left( 
\frac{ 2^{\widetilde{n}_c-1+n_f} \widetilde{\La}_{n_f-n_c-2,n_f}^
{3\left(\widetilde{n}_c+1\right)-n_f}  \mbox{Pf} M }
{ \kappa^{n_f}} \right)^{1/\left(\widetilde{n}_c+1\right)}+W_{tree}(M)
\nonu \\ 
& = &  \left(-n_c +n_f-1 \right)
\ep_{-n_c+n_f-1}
\left( 
\frac{ 2^{-n_c+2n_f-3} 
\widetilde{\La}_{n_f-n_c-2,n_f}^{3\left(n_f-n_c-1\right)-n_f}
\mbox{Pf} M}
{ \kappa^{n_f} }\right)^{1/\left(-n_c+n_f-1\right)}+W_{tree}(M)
\nonu \\
& = & 
  \left(n_c +1-n_f  \right) \ep_{n_c+1-n_f} \left( 
\frac{2^{n_c-1}  \La_{n_c,n_f}^{3\left(n_c+1\right)-n_f}}
{\mbox{Pf} M}  \right)^
{1/\left(n_c+1-n_f\right)}+W_{tree}(M)
\nonu \\
& = & 
W_{ADS}(M)+W_{tree}(M)
\label{W222}
\eea
which is exactly the same as the one in (\ref{eW}). 
Here we used the fact that 
$\widetilde{N}_c = N_f -N_c -4$ and
$\ep_{
\widetilde{n}_c+1}=\ep_{-n_c+n_f-1}=\ep_{n_c-n_f+1}$.
This suggests Seiberg duality in the context of matrix model.
The scale of the magnetic theory in the gauge theory side was 
related to that of 
electric theory by \cite{ip}
\bea
 \La_{n_c,n_f}^{3\left(n_c+1\right)-n_f} 
\widetilde{\La}_{n_f-n_c-2,n_f}^{3\left(n_f-n_c-1\right)-n_f} = 
2^{4-2n_f} \left(-1\right)^{n_f-n_c-1} 
\kappa^{n_f}
\label{scale}
\eea
where the normalization factor  was chosen to get
the consistent low energy behavior under large mass deformation and along the 
flat directions. Note that we take different convention for mass term and that 
is the reason why there exists $2^{-2n_f}$ factor in the above (\ref{scale}).   
Of course, the number of vacua can be also obtained from 
the vacuum  equation in the magnetic theory like as the electric theory.
By making the ansatz 
$M=
i \si_2 \otimes \mbox{diag}(\la_1, \cdots, 
\la_{n_f})$, we get similar vacuum equation. In the massless limit,
the solution for $X$ will provide \cite{carlinoetal1} 
\bea
X^{2n_c+2-n_f} = \widetilde{\La}^{-2(3(n_f-n_c-1)-n_f)} \kappa^{2n_f} 
\mu^{-n_f}
\nonu
\eea
and together with $2^{n_f-1}$ possibilities coming from the sign choices
for each $\la_i$, it will be the same number as $(2n_c-n_f+2) 2^{n_f-1}$
as we have seen before.


\vspace{1cm}
\centerline{\bf Acknowledgments}

This research was supported by 
Korea Research Foundation Grant(KRF-2002-015-CS0006).        
I would like to thank K. Ohta for the correspondence on his paper and 
S. Nam for relevant discussions.

\end{document}